\documentclass{iau}
\usepackage{graphicx}
\usepackage{natbib}
\usepackage{amssymb}
\usepackage{lipsum}

\newcommand{\kepler}{\textit{Kepler}}

\title[
Detecting activity cycles of late-type dwarfs in Kepler data
] 
{Detecting activity cycles of late-type dwarfs in Kepler data}

\author[Kriszti\'an Vida \and Katalin Ol\'ah]   
{Kriszti\'an Vida$^1$
\and Katalin Ol\'ah$^1$}

\affiliation{$^1$Konkoly Observatory of the Hungarian Academy of Sciences, \\
H-1121 Budapest, Konkoly Thege M. str. 15--17. Hungary \\ email: {\tt vidakris@konkoly.hu},
\tt{olah@konkoly.hu}
}

\pubyear{2008}
\volume{xxx}  
\pagerange{119--126}
\jname{Magnetic Fields Throughout Stellar Evolution}
\editors{A.C. Editor, B.D. Editor \& C.E. Editor, eds.}
\begin{document}

\maketitle

\begin{abstract}
Using data of fast-rotating active dwarf stars in the Kepler database, we perform time-frequency analysis of the light curves in order to search for signs of activity cycles. We use the phenomenon that the  active region latitudes vary with the cycle (like the solar butterfly diagram), which causes the observed rotation period to change as a consequence of differential rotation. We find cycles in 8 cases of the 39 promising targets with periods between of 300--900 days.

\end{abstract}

\firstsection 
\section{Introduction}

Activity cycles on the Sun and other active stars are known for a long time. The difficulty of searching  such cycles is that one needs long-term measurements from a few years to decades, as these are  the typical time scales of the phenomenon. For that reason, long-term all-sky surveys  seem a good option in the future, given they run for many years. Now, however, the \kepler{} satellite observes the most precise quasi-continuous light curves of about 150\,000 targets, operating for a few years.
\cite{1996ApJ...460..848B}, and later \cite{rot-cyc} showed that the cycle length depends on rotation: cycles on fast-rotating stars are known to be shorter. 
 \cite{shortcyc} found, that on ultrafast-rotating active stars it is possible to detect activity cycles with lengths of about a year, already within the reach of \kepler. 

The light curves of \kepler{} are unfortunately not free from instrumental, long-term trends and shorter glitches, and although there are methods intended to correct for these, getting a homogeneous trustworthy dataset spanning for multiple quarters seems very hard, if even possible. Thus, when searching for activity cycles in spottedness simply using photometry, these trends interfere with the actual signal we are looking for. 

Beside the change of spottedness, stellar activity cycles have another property -- the change of latitude over the cycle, which effect is known on the Sun as the "butterfly diagram". \cite{emre} showed, that the shape of the butterfly diagram changes fundamentally with the spectral type and rotation period, but on fast-rotating late-type stars the features corresponding to the "wings" of the solar butterfly diagram might be still distinguishable. This effect, together with the differential rotation of the stellar surface (i.e., that the equatorial region rotates faster than the poles), can help us to detect activity cycles. From the butterfly diagram we can learn, that the typical latitude of the active regions change with the cycle: in the case of the Sun, at maximum activity the spots appear up to $30^\circ$ latitude, while at minimum they appear close to the equator.  As a consequence, the differential rotation of the surface causes the typical observable rotation period (periodic light curve modulation as a result of spottedness) to change with the activity cycle. This effect -- the observed rotational period -- is not influenced by the long-term trends of the light curves, thus we can use it to detect the cycles also in \kepler{} data.

\section{Data and methods}
To find objects with sufficiently short activity cycles, similar to the ones in \cite{shortcyc}, we searched the Kepler Input Catalogue (KIC) for dwarfs ($\log g \approx 4.5$) cooler than 4500\,K. We automatically analyzed the Q1 light curves of these 8826 objects to find the main period in the data using the discrete Fourier-transformation option of \textit{TiFrAn} (Time-Frequency Analysis package, see \citealt{tifran} and \citealt{2009AA...501..695K}). Supposing that the largest peak in the Fourier-spectrum  corresponds to the rotation periods of the stars, we inspected visually all the short-period ($P_\mathrm{rot}\lesssim 1 d$) targets to select 39 objects, where the changes in the light curves indicated spottedness. For these targets, all the available data were downloaded (until Q13), and the long-term instrumental changes were removed from the light curves by fitting a third-order polynomial to each quarter.

The resulting light curves were cleaned from extremely outlying points (possibly flares), and were analyzed using the Short-Term Fourier-Transform (STFT) option in \textit{TiFrAn} to find promising targets with possible activity cycles -- i.e., where the STFT shows periodic changes. 
\section{Results and discussion}
\begin{figure}
\centering
\includegraphics[width=0.32\textwidth]{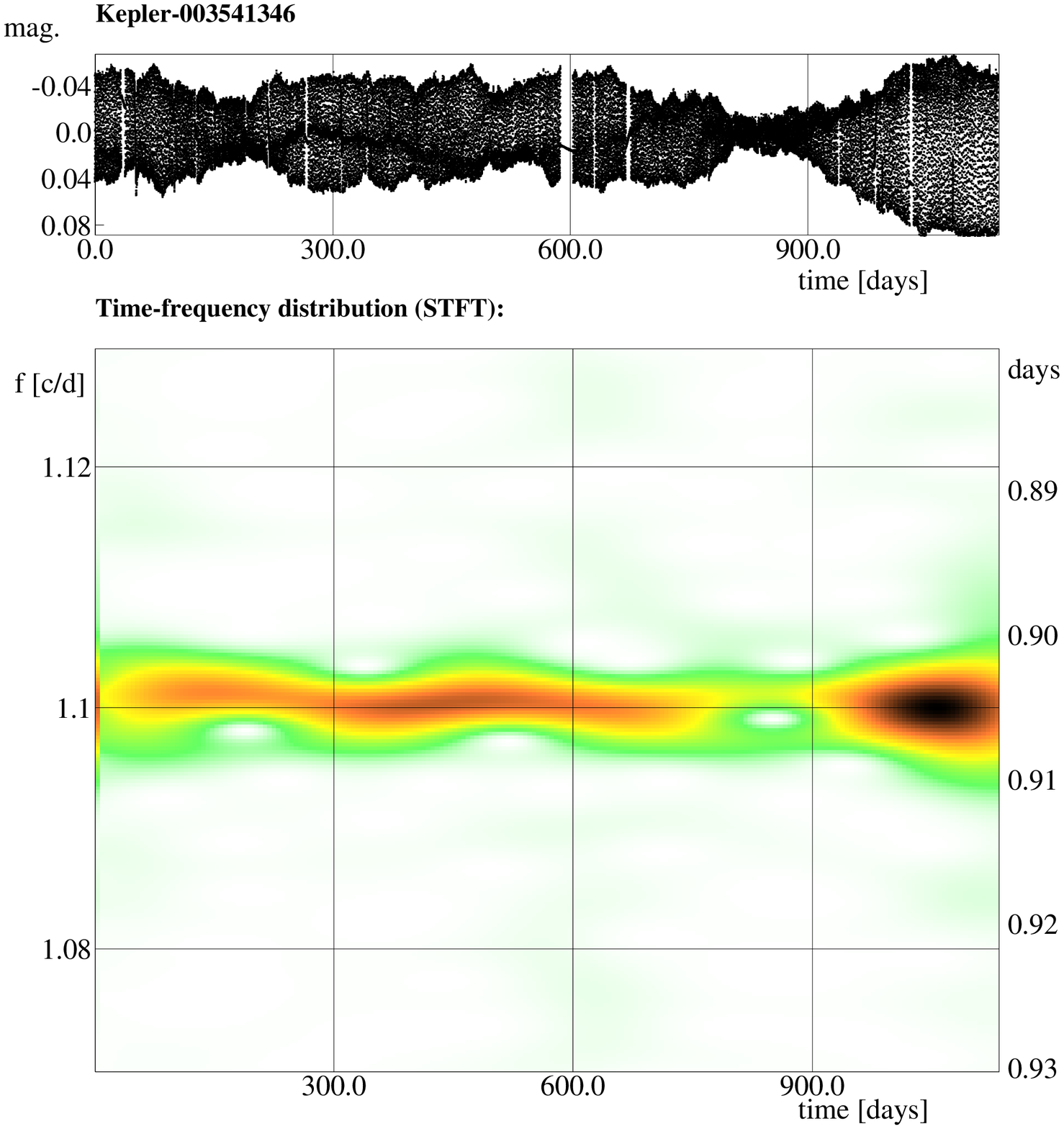}
\includegraphics[width=0.32\textwidth]{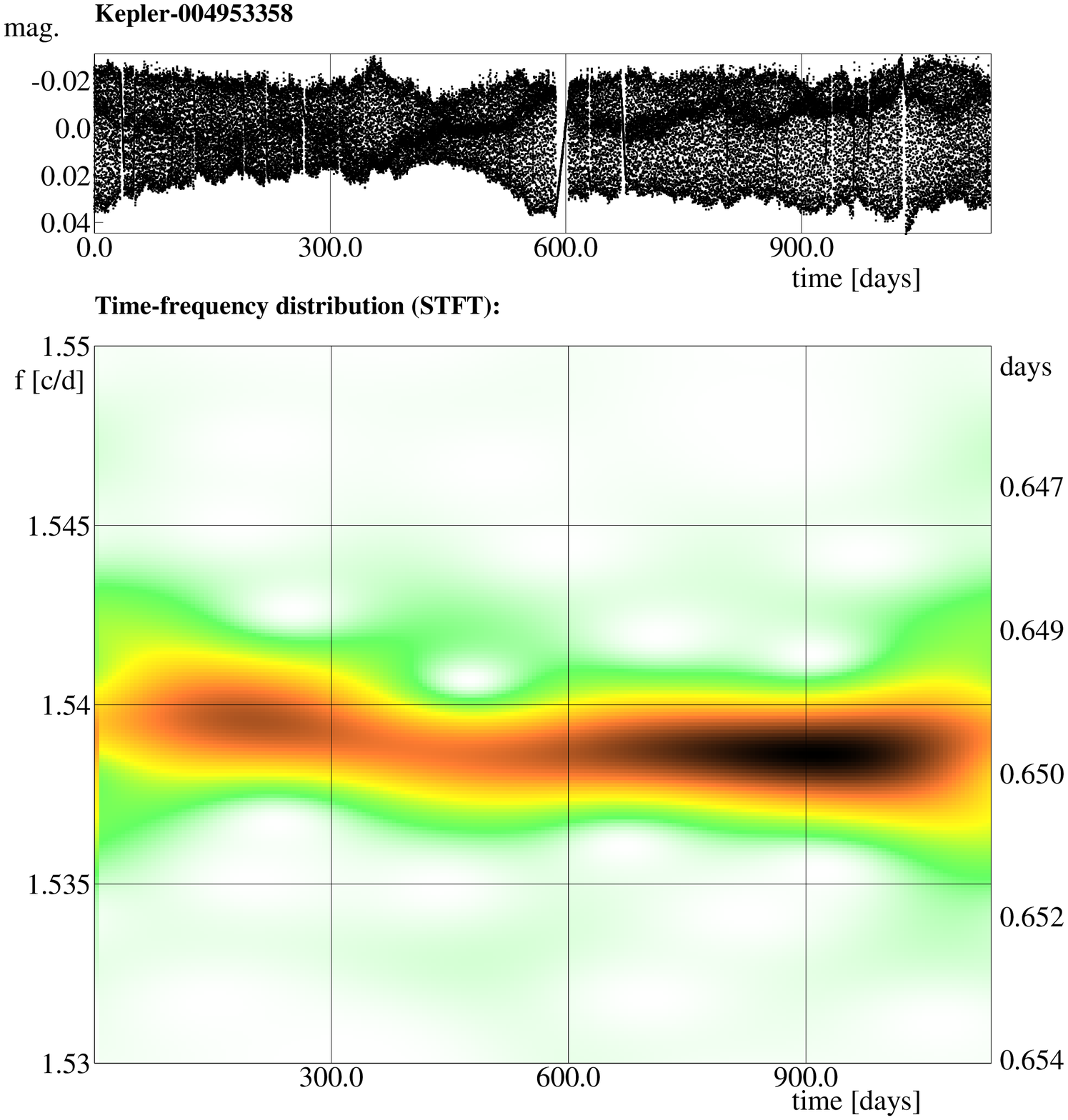}
\includegraphics[width=0.32\textwidth]{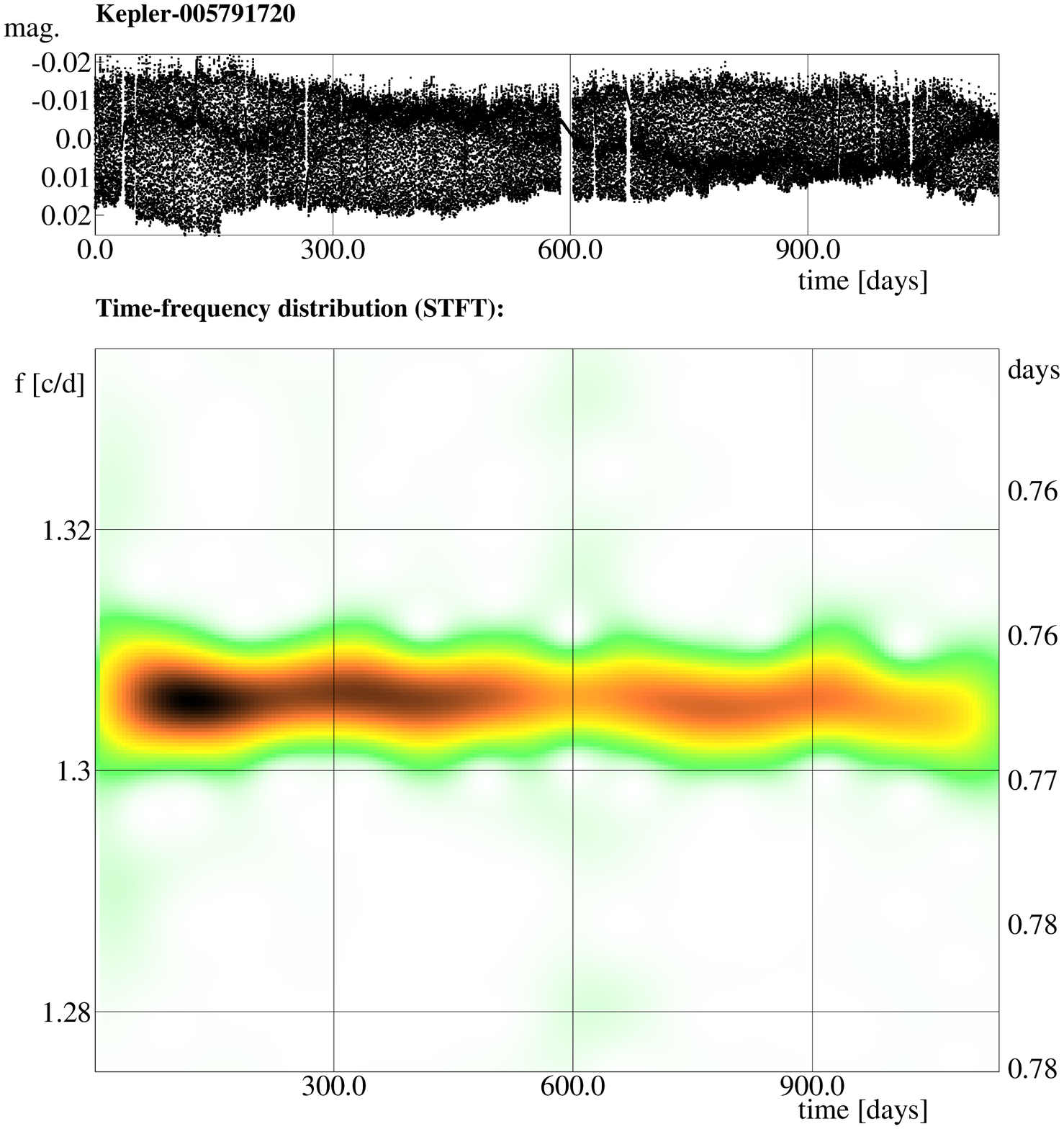}
\includegraphics[width=0.32\textwidth]{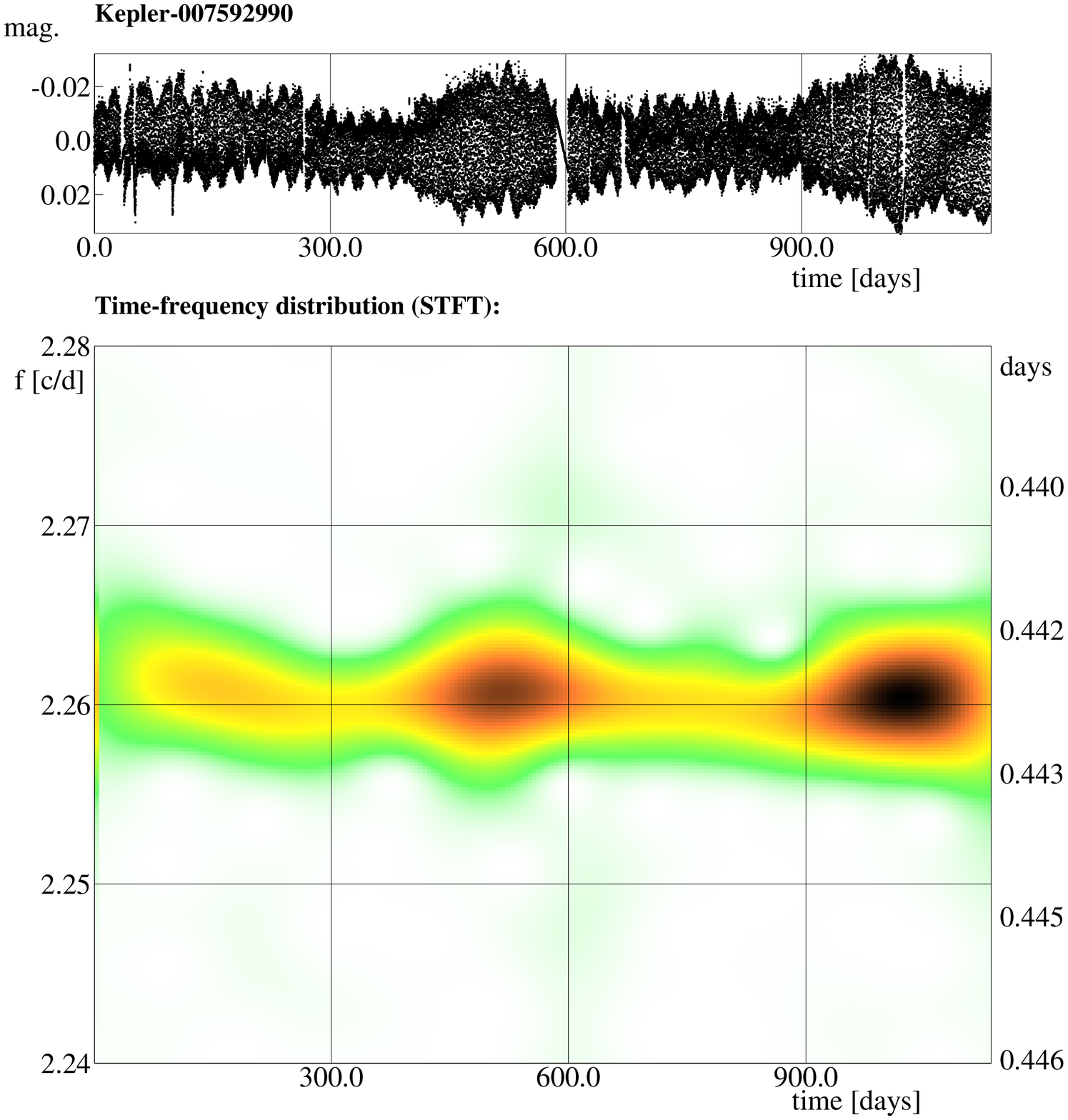}
\includegraphics[width=0.32\textwidth]{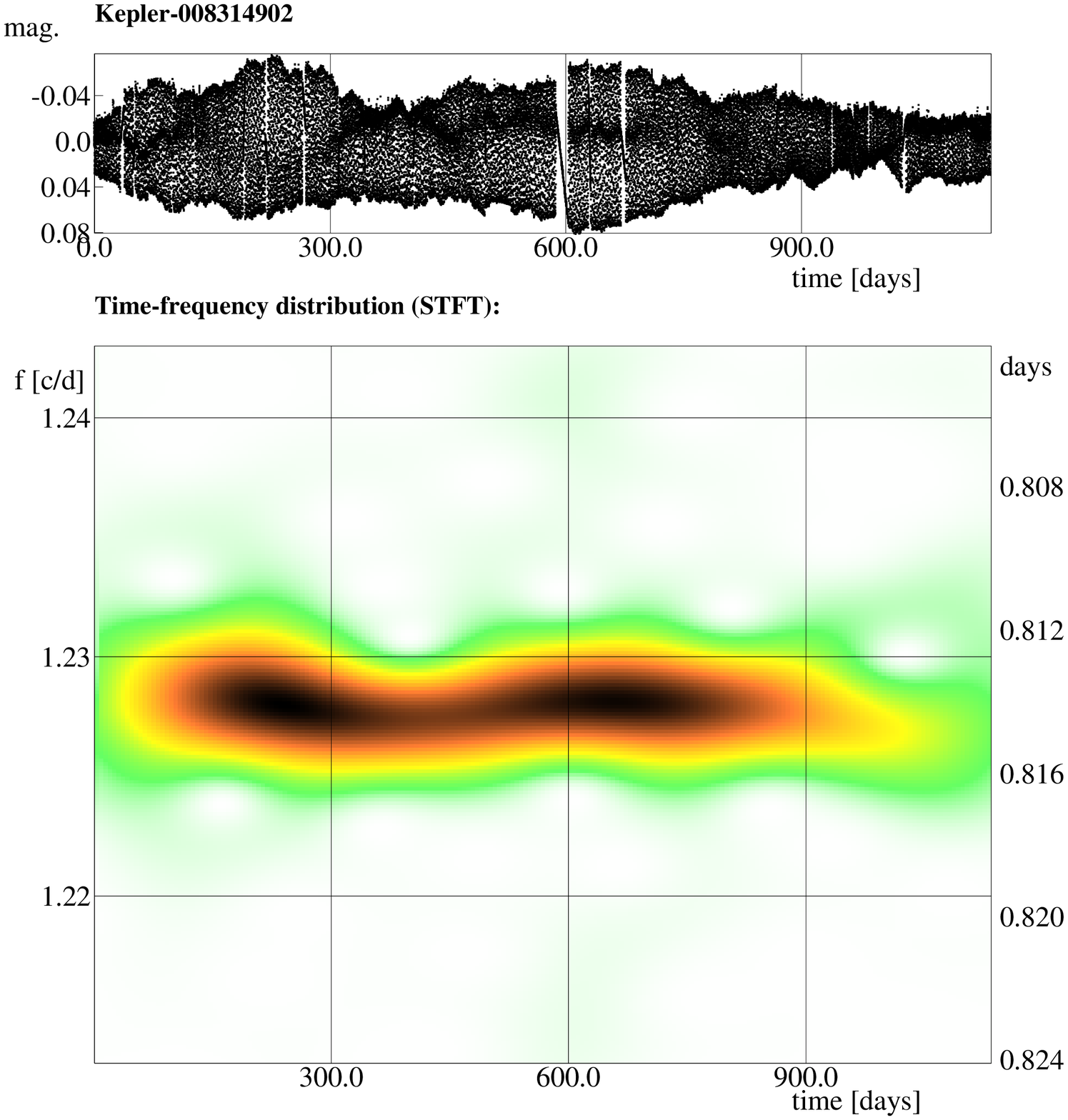}
\includegraphics[width=0.32\textwidth]{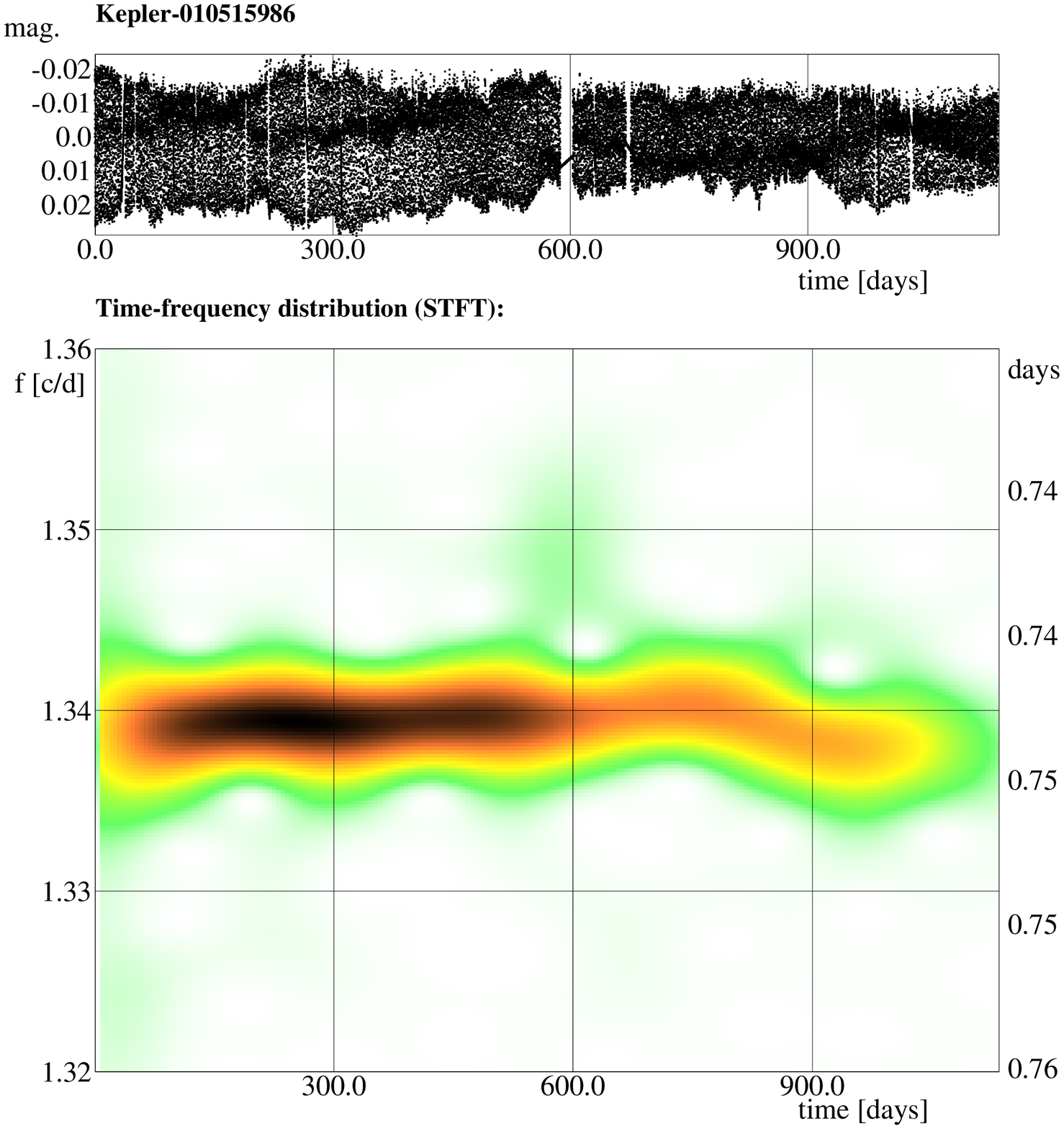}
\includegraphics[width=0.32\textwidth]{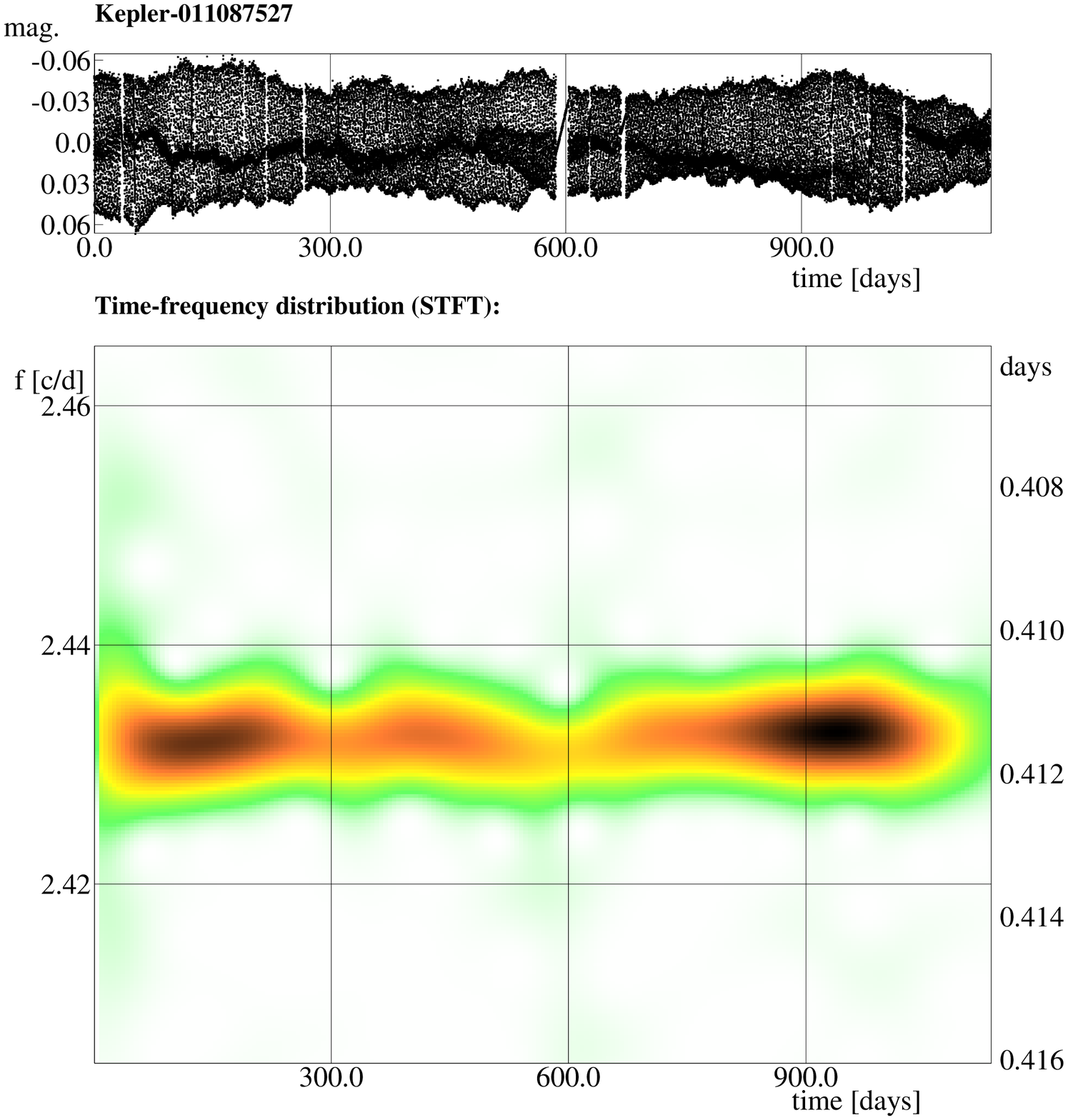}
\includegraphics[width=0.32\textwidth]{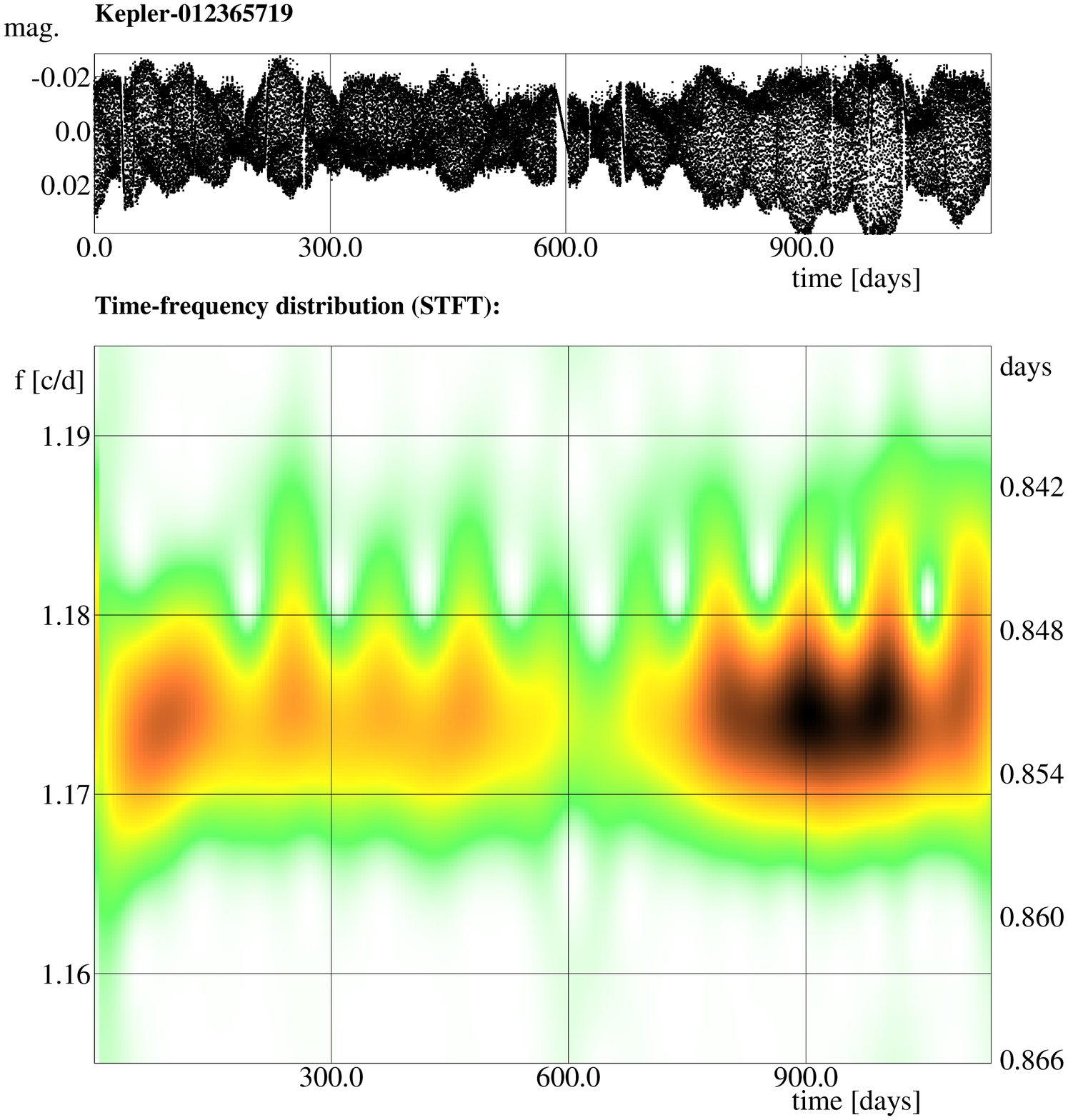}
\includegraphics[width=0.32\textwidth]{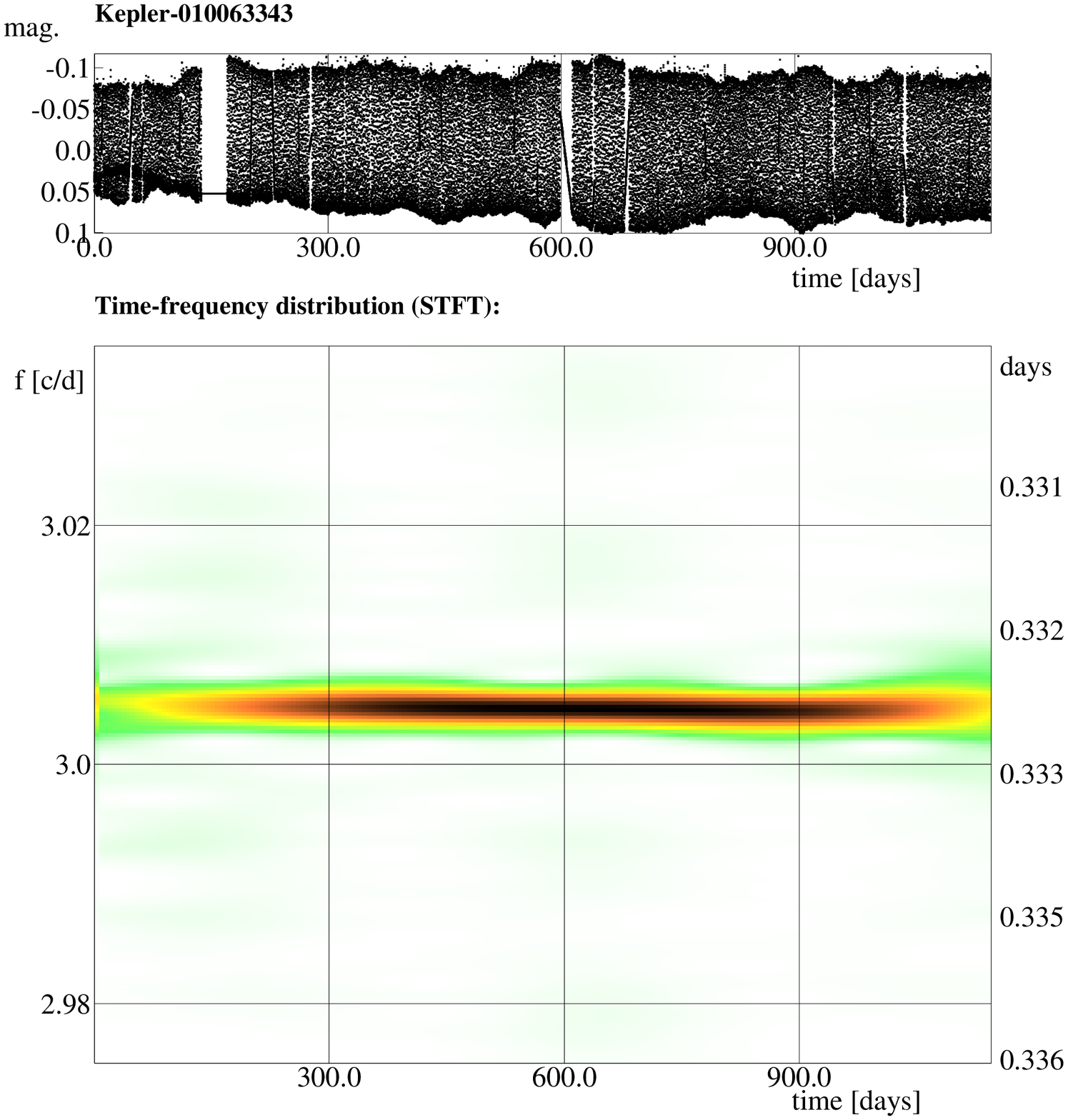}
\caption{Light curves and their Short-Term Fourier-Transforms of the 8 \kepler -targets that possibly have an activity cycle in a range of 400--900 days. The last plot shows a comparison, where no sign of cyclic variation has been found.}
\label{fig:stft}
\end{figure}

\begin{table}
\centering
\caption{Basic parameters of the \kepler -targets with detected activity cycles, their determined rotation period, and the cycle period. The "L" sign denotes long-term changes with the possible length of the cycle shown in parentheses. Last line shows the comparison star without detected cyclic changes.}
\label{tab}
\begin{tabular}{ccccccc}
KeplerID & Kp. mag. &$T_\mathrm{eff}$ & $\log g$ & $P_\mathrm{rot}$ (d) & $P_\mathrm{cyc}$ (d) \\
\hline
03541346 & 15.379 & 4194 & 4.503 & 0.9082& 500  \\ 
04953358 & 15.487 & 3843 & 4.608 & 0.6490& L (800) \\
05791720 & 14.067 & 3533 & 4.132 & 0.7651& 320 \\
07592990 & 15.788 & 4004 & 4.632 & 0.4421& 480\\
08314902 & 15.745 & 4176 & 4.480 & 0.8135  & 580 \\
10515986 & 15.592 & 3668 & 4.297 & 0.7462& L (780)\\
11087527 & 15.603 & 4303 & 4.556 & 0.4110& 300\\ 
12365719 & 15.843 & 3735 & 4.473 & 0.8501& L (820)\\
\hline
10063343 & 13.164 & 3976 & 4.433 & 0.3326& -- \\
\end{tabular}
\end{table}

\begin{figure}
\centering
\includegraphics[width=0.7\textwidth]{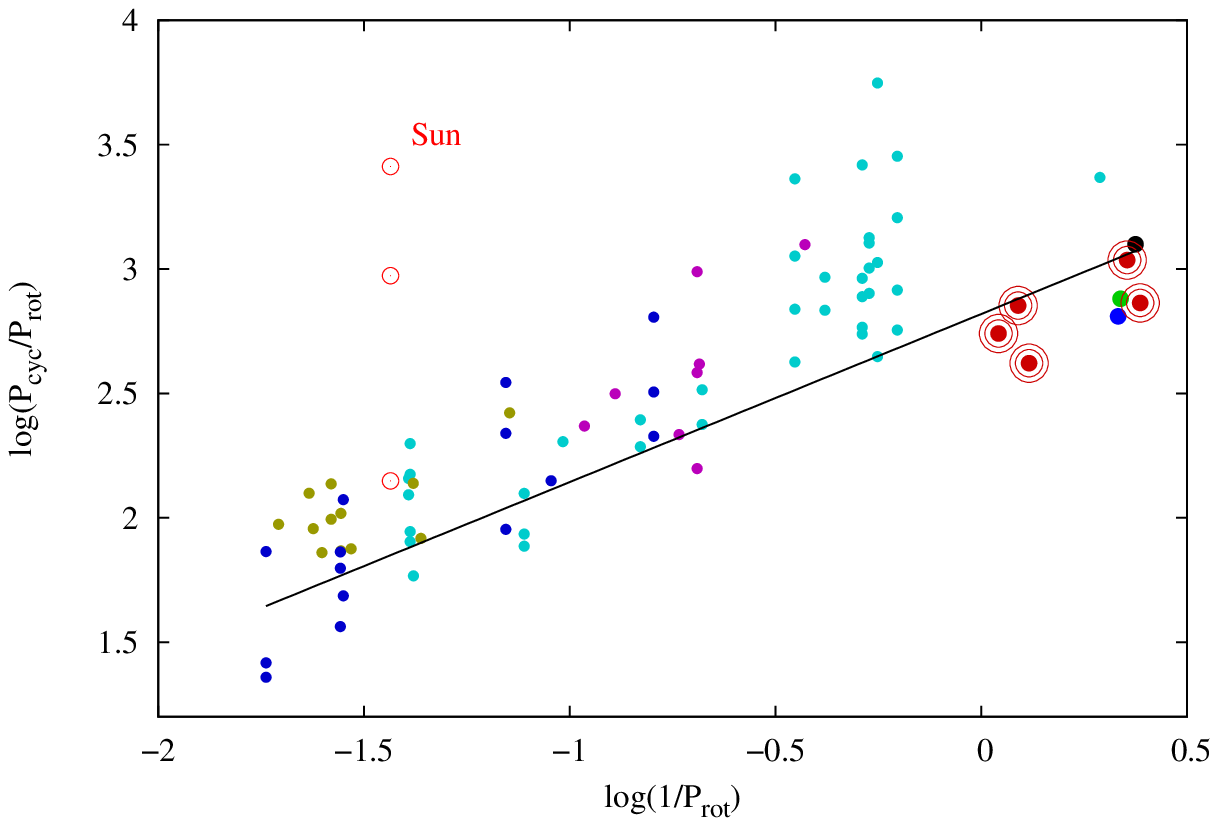}
\caption{Relation between the rotation period and cycle lengths, bases on the plot of \cite{rot-cyc}. Different colors denote different surveys, the line indicates the fit to the shortest cycle lengths. Encircled points show targets from this paper, where the cycle length was short enough to allow a reasonable estimation.}
\label{fig:rot-cyc}
\end{figure}

Of the 39 promising cool dwarf stars with fast rotation, we found signs of an activity cycle in 8 cases. The STFTs and light curves of these targets are shown in Fig. \ref{fig:stft}. The lengths of the cycles are between 300--900 days (the rotation and cycle periods, and basic stellar parameters are summarized in Table \ref{tab}). The cycle lengths were determined both by visual inspection and using 1D cross-correlation and Fourier-analysis of the STFSs to give a quantitative result (Table \ref{tab} shows the latter values).

Five stars show definite cycles with lengths between 300--580 days (1--1.5 years). The remaining three, denoted with "L", show also systematic period change, but the time-base, about 4 years, is not long enough to find periods with great confidence over 2 years.

In Fig. \ref{fig:rot-cyc} we plotted our results on the rotation period--cycle length diagram based on \cite{rot-cyc}, showing only those 5 targets, where the length of the cycle is short enough compared to the length of the dataset to reasonably estimate the cycle length. 
The stars with definite cycles of about 1--1.5 years are quite similar to the original sample from \cite{shortcyc}, on which the target search was based on, and fill in the currently unmapped part of the rotation--cycle length diagram.
In that paper the following activity cycles were reported: 
EY\, Dra              ($P_\mathrm{rot}=0.459d$, $P_\mathrm{cyc}=348$d),
V405\,And           ($P_\mathrm{rot}=0.465d$, $P_\mathrm{cyc}=305$d), and
GSC\, 3377-0296 ($P_\mathrm{rot}=0.445d$, $P_\mathrm{cyc}=530$d).
Using the activity cycles found in the present paper and those from \cite{shortcyc}, we can refine the correlation (see Fig. \ref{fig:rot-cyc}). With the targets of the present paper, where the activity cycles were short enough to give a fair estimate for their lengths, for the shortest cycles (several objects, including the Sun show multiple cycles), we get the following relation:
$$\log (P_\mathrm{cyc}/P_\mathrm{rot} )=0.68 \log(1/P_\mathrm{rot})+2.82$$
 for the sample of both single and binary stars from different surveys. \cite{shortcyc} found, that the lengths of the activity cycles for ultrafast-rotating dwarfs seems to be somewhat shorter than the previous samples would indicate. 
 The objects from \cite{shortcyc} served as template to search for interesting \textit{Kepler} targets, those results were the base of the present study.

At present we have altogether 8 stars with rotational periods between 0.4--1.0 days showing activity cycles of 1--1.5 years. As seen from Fig. \ref{fig:rot-cyc}, these objects continue the trend showing shorter cycle lengths when rotating faster.

\acknowledgement
The financial support of the OTKA grant K-81421, is acknowledged.
This work was also supported by the ``Lend\"ulet-2009'', and ``Lend\"ulet-2012'' Young Researchers' Programs of the Hungarian Academy of Sciences, and by the HUMAN MB08C 81013 grant of the MAG Zrt.
Funding for the \textit{Kepler} mission is provided by NASA's 
Science Mission Directorate.

\begin{discussion}

\end{discussion}

\end{document}